# TITAN'S ORGANIC AEROSOLS: MOLECULAR COMPOSITION AND STRUCTURE OF LABORATORY ANALOGUES INFERRED FROM PYROLYSIS GAS CHROMATOGRAPHY MASS SPECTROMETRY ANALYSIS

Marietta Morisson[a,b,*], Cyril Szopa[b,c], Nathalie Carrasco[b,c], Arnaud Buch[a] and Thomas Gautier[d]

[a] LGPM Laboratoire de Génie des Procédés et Matériaux, Ecole CentraleSupelec, Châtenay-Malabry Cedex, France
[b] LATMOS/IPSL, UVSQ Université Paris-Saclay, UPMC Univ. Paris 06, CNRS, Guyancourt, France
[c] Institut Universitaire de France, Paris F-75005, France
[d] NASA Goddard Space Flight Center, Greenbelt, MD, United States

**Abstract**

Analogues of Titan's aerosols are of primary interest in the understanding of Titan's atmospheric chemistry and climate, and in the development of *in situ* instrumentation for future space missions. Numerous studies have been carried out to characterize laboratory analogues of Titan aerosols (tholins), but their molecular composition and structure are still poorly known. If pyrolysis gas chromatography mass spectrometry (pyr-GCMS) has been used for years to give clues about their chemical composition, highly disparate results were obtained with this technique. They can be attributed to the variety of analytical conditions used for pyr-GCMS analyses, and/or to differences in the nature of the analogues analyzed, that were produced with different laboratory set-ups under various operating conditions.

In order to have a better description of Titan's tholin's molecular composition by pyr-GCMS, we carried out a systematic study with two major objectives: (i) exploring the pyr-GCMS analytical parameters to find the optimal ones for the detection of a wide range of chemical products allowing a characterization of the tholins composition as comprehensive as possible, and (ii) highlighting the role of the $CH_4$ ratio in the gaseous reactive medium on the tholin's molecular structure. We used a radio-frequency plasma discharge to synthetize tholins with different concentrations of $CH_4$ diluted in $N_2$. The samples were pyrolyzed at temperatures covering the 200-700°C range. The extracted gases were then analyzed by GCMS for their molecular identification.

The optimal pyrolysis temperature for characterizing the molecular composition of our tholins by GCMS analysis is found to be 600°C. This temperature choice results from the best compromise between the number of compounds released, the quality of the signal and the appearance of pyrolysis artifacts. About a hundred molecules are identified as pyrolysates. A common major chromatographic pattern appears clearly for all the samples even if the number of released compounds can significantly differ. The hydrocarbon chain content increases in tholins when the $CH_4$ ratio increases. A semi-quantitative study of the nitriles (most abundant chemical family in our chromatograms) released during the pyrolysis shows the existence of a correlation between the amount of a nitrile released and its molecular mass, similarly to the previous quantification of nitriles in the plasma gas-phase. Moreover, numerous nitriles are present both in tholins and in the gas phase, confirming their suspected role in the gas phase as precursors of the solid organic particles.

## I Introduction

Titan's surface is well known for being hidden by a thick photochemical haze made of organic aerosols. Knowledge of the aerosol's chemical composition is of major importance since they play a prominent role in the radiative equilibrium of the satellite (Flasar et al., 2005; McKay et al., 1991). They contribute to the surface spectral signature (Tomasko et al., 2005), act as a sink for carbon-containing molecules in the gas phase (Lebonnois, 2002) and probably as condensation nuclei for the formation of clouds (Mayo and Samuelson, 2005). Moreover, Titan is the only body in the Solar System – besides Earth – to have a well-established complex organic chemistry. This particularity makes Titan one of the most interesting objects for astrobiology and the search for an extraterrestrial form of prebiotic chemistry in the Solar System (Raulin, 2007).

The chemical composition of Titan's organic aerosols remains poorly characterized in spite of the numerous results from the Cassini-Huygens space mission. The Cassini/VIMS and Cassini/CIRS infrared spectrometers did not provide detailed data on the aerosol composition (Hirtzig et al., 2013; Vinatier et al., 2012), neither did the Huygens/DISR experiment which analyzed the surface and atmosphere during the Huygens probe descent in 2005 (Tomasko et al., 2005). The Huygens/ACP-GCMS experiment analyzed the chemical composition of aerosols collected in the atmosphere. The condensed molecules present in the samples were released during a first heating at 250°C and analyzed by Gas Chromatography Mass Spectrometry. The same procedure was done at 600°C to characterize the products of pyrolysis of the refractory core. Technical issues encountered by the ACP experiment were reported in the supplementary information of the Israel et al. (2005) article. They involved a leak of the sealing valve of the oven. The lack of tightness during the analysis prevented the efficient transfer of the volatile molecules released at 250°C to the GCMS experiment, resulting in the absence of an obvious GCMS signature of these molecules. These ACP issues however did not affect the MS analysis of the remaining refractory part of the sample extracted at 600°C. The experiment detected hydrogen cyanide (HCN) and ammonia ($NH_3$) in the 600°C pyrolysis products, proving that the aerosols are made up of a core of refractory organics - including N-bearing compounds - without any other information about the aerosol's molecular structure (Israel et al., 2002; Israël et al., 2005). Laboratory experiments are thus, to date, more likely to provide information about physical and chemical properties of the aerosols. This is the reason why Titan's aerosol laboratory analogues, named "tholins" in the following, have been studied since the 1980's (Khare et al., 1981).

Pyrolysis-GC-MS is one of the most commonly used techniques for the molecular analysis of tholins, because of its efficiency and simplicity of implementation. However the variability in the nature of the analogues (Cable et al., 2012), as well as in the conditions, instruments and techniques used for the analyses, leads to disparate results which are difficult to compare. Table 1 lists the analytical and sample synthesis conditions for all the previous pyr-GCMS studies. It shows that at least one parameter changes from one study to another, preventing definitive comparisons. For example, the choice of the chromatographic column impacts the type of compounds that can be analyzed. For instance, the use of a porous layer open tubular (PLOT) column generally focuses on volatile compounds ($C_1$ to $C_5$), whereas wall coated open tubular (WCOT) columns rather allow the separation of heavier compounds.

The most recent tholins analyses dismiss the hypotheses of a purely polymeric (poly-HCN, poly-$HC_3N$) or co-oligomeric (HCN-$C_2H_2$ or HCN- $HC_3N$) structure (Israel et al., 2002; Israël et al., 2005; Khare et al., 1981), in favor of a more irregular structure (Coll et al., 2013). Studies carried out with *a priori* comparable samples – synthetized from equivalent gas mixtures ($N_2$:$CH_4$, ratio 90:10) with cold plasma discharges – still diverge about the nature of the macromolecules. They are mostly aromatic or poly-aromatic hydrocarbons according to the Pyr-GCMS analysis by (Coll et al., 1998). The study by (McGuigan et al., 2006) reveals N-heterocycles (pyrroles) by GCxGC-TOF-MS analysis. And the nature of the macromolecules is found to be an open-chain structure according to the TG-MS study by (Nna-Mvondo et al., 2013), who, on the contrary, do not detect any aromatic, polyaromatic or cyclic compound.

The compounds detected in previous studies of thermal degradation of tholins are listed in table 2. All these molecules, produced from tholins pyrolysis, are divided into seven main chemical families: alkanes, alkenes, alkynes, aliphatic nitriles, aromatic nitriles, aromatic and cyclic hydrocarbons, and nitrogenous heterocycles. The "isomer" section of the table gives the empirical formula of the compounds whose exact isomeric structure could not be determined.

We should also mention that – except in the case of the study done by (Khare et al., 1981), where water was used in the initial gaseous mixture – the presence of oxygen bearing molecules, listed in the last section of the table, is due to a contamination, most probably by the molecular oxygen from the ambient air (Brassé, 2014).

As reported in table 1, thermal degradation studies have been carried out on tholins produced under various experimental conditions. One of the main parameters is the composition of the initial gas mixture used for the synthesis (only $CH_4$ and $N_2$, or in presence of $H_2O$ or $H_2$), but the effect of the pressure inside the reaction chamber is also investigated in (Imanaka et al., 2004). In the case of $N_2$:$CH_4$ mixtures, the 90:10 ratio is widely favored. Samples produced from mixtures with different $CH_4/N_2$ ratios are analyzed by GC-MS after a 750°C pyrolysis by (Coll et al., 2013). They however do not find any qualitative difference among the pyrolysates. They also compare the Huygens ACP experiment data with their results – obtained from pyrolysis of four kinds of samples: hydrogen cyanide polymer, solid hydrocarbons (polyethylene and anthracene), and tholins synthetized with cold and hot plasmas. The plasma is "cold" when the plasma heating does not significantly alter the temperature of the neutral gas. Coll et al. conclude that tholins produced in a cold plasma are the most similar analogues to Titan's aerosols regarding the volatiles released during pyrolysis.

Analysis conditions represent an additional variable, in terms of choice of chromatographic column (*e.g.* CPSil5-CB (Coll et al., 1998), DB-1 and RTX-Wax (McGuigan et al., 2006), PoraPLOT Q (Buch et al., 2006; Coll et al., 2013; Pietrogrand et al., 2001) and DB-5 (Ehrenfreund et al., 1995; Khare et al., 1984)) as well as of column temperature program, pressure conditions or pyrolysis temperature. These parameters are decisive for the kind of molecules, which are analytically detectable (polar/non-polar, heavy/light) with GCMS analysis. Lastly, the choice of pyrolyzer is important in the thermal decomposition process from solid to gas phase. Some pyrolyzers are indeed limited in the reachable pyrolysis temperature, as are the Curie point pyrolyzers. The geometry of the pyrolyzers and the residence-time of the pyrolysates are also variable, leading to a different recombination of species inside the oven (Moldoveanu, 1998). All these varying parameters explain the heterogeneity of the published studies.

| N° | Publication | Gas mixture | Energy source | Analysis method | Chromatographic column | $T_{pyrolysis}$ | Column Pressure/ Flow | GC temperature program |
|---|---|---|---|---|---|---|---|---|
| 1 | Khare & Sagan, 1981 (Khare et al., 1981) | $CH_4/NH_3/H_2O$ (51.5 : 45.9 : 2.6) | Hot plasma (spark discharge) | Pyr-GCMS | 150' × 0.02"i.d. OS 138-polyphenylether/SCOT | 150°C- 600°C (c) 600°C (s) | N.D. | 10 min at 40°C<br>2.5°C.min$^{-1}$ to 190°C |
| 2 | Khare et al., 1984 (Khare et al., 1984) | $N_2/CH_4$ (90 : 10) | Cold plasma (direct current electrical discharge) | Pyr-GCMS | DB-5 fused silica 30m | Room T - 700°C (c) and (s) | 2 cc.min$^{-1}$ | 4°C.min$^{-1}$ from room temperature to 250°C |
| 3 | Scattergood, 1987 (Scattergood, 1987) | $N_2/CH_4/H_2$ (96.8 : 3 : 0.2) | Hot plasma (spark discharge) | Pyr-GC | Porapak | 20°C-700°C (s) | N.D. | N.D. |
| 4 | Ehrenfreund et al., 1995 (Ehrenfreund et al., 1995) | $N_2/CH_4$ (90 : 10) | Cold plasma (corona discharge) | Pyr(Curie)-GCMS | J & W DB-5 fused silica 30m, 0.25 mm ID | 770°C | N.D. | 4°C.min$^{-1}$ from -10°C to 300°C |
| 5 | Clarke & Ferris, 1997 (Clarke and Ferris, 1997) | (poly-$HC_3N$) | UV (185 nm) | Pyr-RMN/IR | - | - | - | - |
| 6 | Coll et al., 1998 (Coll et al., 1998) | $N_2/CH_4$ (90 : 10) | Hot plasma (spark discharge) | Pyr(oven)-GCMS | CPSil 5 CB 25 m, 0.15 mm ID | 600°C | 1,6 bar | 20 min at 30°C<br>4°C.min$^{-1}$ to 150°C<br>10 min à 150°C |
| 7 | Pietrogrande et al., 2001 (Pietrogrand et al., 2001) | $CH_4/N_2$ (90 : 10) | Cold plasma (corona discharge) | Pyr(Curie)-GCMS | PoraPlot Q fused-silica 25 m, 0.32mm ID | 750°C | N.D. | 2 min at 60°C<br>10°C.min$^{-1}$ to 240°C<br>30 min at 240°C |
| 8 | McGuigan et al., 2006 (McGuigan et al., 2006) | $N_2/CH_4$ (90 : 10) | Cold plasma discharge | Pyr(Curie) GCxGC-TOF-MS | (1) J & W DB-1 30m, 0.25mm ID<br>(2) Restek Rtx Wax 2m, 0.10mm ID | 250°C-900°C (c) and (s) | 1,5 cc.min$^{-1}$ | (1) 5 min at 40°C then 3°C.min$^{-1}$ to 190°C hold 45 min<br>(2) temperature maintained 5°C greater than (1) |
| 9 | Szopa et al., 2006 (Szopa et al., 2006) | $N_2/CH_4$ (90 : 10) | Cold plasma (capacitively coupled plasma, CCP) | Pyr(oven)-GCMS | PoraPlot Q 30m x 0.25mm x 0.10µm | 650°C | 7,5 psi | 2 min at 100°C<br>10°C.min$^{-1}$ to 240°C<br>40 min at 240°C |
| 10 | De La Fuente et al., 2011 (De la Fuente et al., 2011) | (poly-HCN) | - | TG-MS, DTA, DSC | - | Room T - 1000°C (c) | - | - |
| 11 | De La Fuente et al., 2012 (la Fuente et al., 2012) | $N_2/CH_4/H_2$ (96.8 : 3 : 0.2) | Hot plasma (spark discharge) | TG-MS, DTA, DTG | - | Room T - 1000°C (c) | - | - |
| 12 | Nna-Mvondo et al., 2013 (Nna-Mvondo et al., 2013) | $N_2/CH_4$ (90 : 10) | Cold plasma (Inductive-coupled plasma, ICP) | TG-MS, DTA, DTG | - | Room T - 1000°C (c) | - | - |
| 13 | Coll et al., 2013 (Coll et al., 2013) | (poly-HCN) | Anhydrous or in aqueous solution synthesis | Pyr-GCMS (platinum coil filament) | PoraPlot Q 25m, 0.32 mm ID | 750°C | 1,5 ml.min$^{-1}$ | 2 min at 60°C<br>10°C.min$^{-1}$ to 240°C<br>30 min at 240°C |
| 14 | Coll et al., 2013 (Coll et al., 2013) | $N_2/CH_4$ | Hot plasma (spark discharge, laser) | | | | | |
| 15 | Coll et al., 2013 (Coll et al., 2013) | $N_2/CH_4$ | Cold plasma (corona discharge, CCP) | | | | | |
| 16 | Coll et al., 2013 (Coll et al., 2013) | (poly-$C_nH_n$) | - | | | | | |
| 17 | He et al., 2014 (He et al., 2014) | $N_2/CH_4$ (95 : 5) | Cold plasma (CCP) | TG-MS, DSC, DTG | - | - | - | - |

*Table 1 Summary table of the analysis parameters used for previous studies of Titan tholins thermal degradation. If a pressure or a flow is given in the "column pressure/flow" column, the analysis was done at constant controlled pressure or constant controlled flow respectively. NMR = Nuclear Magnetic Resonance; IR = Infrared; TOF-MS = Time Of Flight Mass Spectrometry; TG =*

Considering the disparity in the nature of the samples and in the analytical cond previous studies, we carry out a systematic analysis of tholins produced in well-co GCMS conditions. Besides determining the composition and structure of these analog aerosols, the major objectives of this study are:

(i) To determine the variability in the tholins molecular composition for sam with different initial gas mixtures (methane concentrations between 2% an same conditions and with a given experimental device. The influence of concentration on the tholins chemical composition has been highlighted O'Brien et al., 2010), and (Pernot et al., 2010). (Sciamma-O'Brien et al., 201 elementary analyses on tholins samples synthesized with the PAMPRE 2010. They showed that the C/N ratio increases when the $CH_4$ concentratio (Pernot et al., 2010) a study of the tholins soluble fraction by high-re spectrometry showed an increase in the number of methylene (-$CH_2$-) grou to nitrogen-bearing groups and a higher complexity of the material w concentration increases. The different samples studied here have therefor variability in their chemical composition, but their molecular variation addressed yet;

(ii) To study the influence of the pyr-GCMS analytical conditions on the detected species. The main parameters to be studied are: the influence o temperature on the nature of the released products; the temperature pro heat the GC column in order to find a tradeoff for the analysis of both th the heaviest compounds; the separation of the same pyrolysates by columns.

**AMMONIA**

Ammonia [1, 4, 5, 7, 10, 12, 14, 17]

**ALKANE HYDROCARBON**

| | | |
|---|---|---|
| Methane[3, 5, 7, 9, 11, 12, 13, 15] | Isobutane[2, 3] | Pentadecane[8] |
| Ethane[1, 3, 5, 7, 13, 14, 15, 17] | 2-Methylbutane[7] | Tridecane[8] |
| Propane[2, 3, 12, 13, 15, 17] | Pentane[1, 16] | Dodecane[8] |
| 2-Methylpropane[7, 13] | Hexane[13] | Undecane[8] |
| Butane[1, 3, 13] | Heptane[2, 13] | Decane[8] |

**ALKENE**

| | | |
|---|---|---|
| Ethylene [3, 5, 7, 11, 12, 13, 14, 15, 16, 17] | Pentene [1, 2] | Heptene [2, 6, 13] |
| Propene [1, 2, 7, 9, 11, 13] | 1-Pentene [13] | Heptadiene [13(2 isomers), 15] |
| Propadiene [7] | 3-Methyl-1-pentene [13] | Pentadecene [8] |
| 1,2-Propadiene [13, 15] | Pentadiene [9(2 isomers), 13, 14, 15] | Tridecene [8] |
| 1-Butene [1, 2, 6, 7, 9, 13, 16] | 1,2-Pentadiene [6] | Dodecene [8] |
| 2-Butene (E) [6, 7, 9, 13, 14] | 1,3-Pentadiene [6, 13] | Undecene [8] |
| 2-Butene (Z) [13] | 1,4-Pentadiene [6, 7] | Decene [8] |
| 2-Methylpropene [6] | 2,3-Pentadiene [6, 7] | 1-Hexene [13] |
| Butadiene [1, 7, 9] | Hexene [2] | 2-Hexene [13] |
| 1,2-Butadiene [13, 15] | 1,5-Hexadiene [13] | 3-Methyl-1-hexene [6] |
| 1,3-Butadiene [6, 14] | 1,3,5-Hexatriene [6] | 5-Methyl-1-hexene [6] |
| 2-Methyl-1,3-butadiene [6] | 2-Methyl-1,3,5-hexatriene [13] | Hexadiene [2] |
| 3-Methyl-1,2-butadiene [6] | | |

**ALKYNE**

| | | |
|---|---|---|
| Acetylene [3, 11, 13, 14, 15] | 1-Penten-3-yne [6] | 2-hexen-4-yne [6] |
| Methylacetylene [3] | 3-Methyl-2-penten-4-yne [6] | 1,3-hexadien-5-yne [6] |
| Propyne [7, 13, 15] | 2-Penten-5-yne [6] | 1,5-hexadien-3-yne [6] |
| 1-Butyne [6, 7, 13, 16] | 3-Hexen-1-yne [6] | 1,5-hexadiyne [6] |
| 2-Butyne [2] | 1-Hexen-3-yne [6] | 2,4-hexadiyne [6] |
| 2-Methyl-1-buten-3-yne [6, 13] | | |

**ALIPHATIC NITRILE**

| | | |
|---|---|---|
| Hydrogen cyanide [1 - 5, 7, 9 - 12, 14 - 17] | Butyronitrile [4, 6, 8, 9] | 2-Cyano-1-butene [2] |
| Acetonitrile [1 - 9, 11, 12, 14, 15, 16, 17] | Isobutyronitrile [2, 4, 6, 7, 8, 14] | Pentanenitrile [1, 16] |
| Cyanogen [3, 7, 9, 14, 16, 17] | But-2-enenitrile [2, 4] | 2-Pentenenitrile [2,6] |
| Propionitrile [1, 2, 4, 5, 7, 8, 11, 14, 15, 17] | Succinonitrile [1] | 3-Pentenenitrile [8] |
| Acrylonitrile [1, 4, 5, 6, 7, 9, 14, 15, 16, 17] | Butenenitrile [9] | 2,4-Pentadienenitrile [8] |
| Propenenitrile [7] | 2-Butenenitrile [6, 8] | 5-Hexenenitrile [2] |
| Methacrylonitrile [2, 4, 6, 7, 14, 15] | 3-Butenenitrile [2, 4, 6, 7, 8] | Hexanenitrile [8] |
| Ethacrylonitrile [6] | | |

**AROMATIC NITRILE**

| | | |
|---|---|---|
| Benzonitrile [1, 2, 6, 7, 8] | Dicyanobenzene [2] | 1,2,4-Tricyanobenzene [(non volatil) 5] |
| Methyl-benzonitrile [7] | 1,3-Dicyanobenzene [(non volatil) 5] | 1,3,5-Tricyanobenzene [(non volatil) 5] |
| Tolunitrile [1, 2] | 1,4-Dicyanobenzene [(non volatil) 5] | |
| **AROMATIC & CYCLIC HC** | | |
| Cyclopropane [11] | 1,2-Dimethyl-benzene [6, 9] | 2-Methyl-1-ethenylbenzene [6] |
| Methylcyclobutane [13] | 1,3-Dimethyl-benzene [6, 9] | 3-Methyl-1-ethenylbenzene [6] |
| 1,2-Dimethyl-cyclopentane [6] | 1,4-Dimethyl-benzene [6, 9] | 4-Methyl-1-ethenylbenzene [6] |
| 1,3-Dimethyl-cyclopentane [6] | Ethylbenzene [6, 8, 9] | Alpha-propynylbenzene [6] |
| Cyclopentene [2, 13] | Ethenylbenzene [2, 6, 7] | Alpha-beta-propadienylbenzene [6] |
| 1,3-Cyclopentadiene [6, 7, 15] | Diethenylbenzene [2] | 4-Methyl-1-ethynylbenzene [6] |
| 1-Methyl-cyclopentadiene [6] | 1,2,3-Trimethylbenzene [6] | Toluene [1, 2, 4, 6, 7, 8, 9, 13, 14, 15] |
| 5-Methyl-cyclopentadiene [6] | 1,2,4-Trimethylbenzene [6] | C3-Alkylbenzene [1] |
| Methylcyclohexane [13] | 1,3,5-Trimethylbenzene [6] | C2-Alkylindane [1] |
| Cyclohexene [13] | Isopropylbenzene [6] | Indene [2, 8] |
| Cyclohexadiene [13] | 2-Methyl-1-ethylbenzene [6] | Methyl-indene [1, 2, 8] |
| 1,3-Cyclohexadiene [6] | 3-Methyl-1-ethylbenzene [6] | 1,3-Dimethyl-indene [8] |
| 1,4-Cyclohexadiene [6] | 4-Methyl-1-ethylbenzene [6] | 1-Methylene-indene [6] |
| 4-Methyl-cyclohexene [13(2 isomers)] | Methyl-ethenylbenzene [2] | Azulene [6] |
| Cycloheptane [6] | Alpha-methyl-ethenylbenzene [6] | Naphthalene [6, 8] |
| Benzene [2, 4, 6, 7, 8, 9, 11, 13, 14, 15, 16, 17] | Beta-methyl-ethenylbenzene [6] | Methyl-naphthalene [2] |
| Dimethyl-benzene [1, 2] | Beta-propenylbenzene [6] | |
| **NITROGENOUS RINGS** | | |
| Dimethyl-aziridine [2] | Pyridine [2, 4, 7, 8, 14, 15, 16] | Pyrimidine [4, 7] |
| Azetidine [2] | Methyl-pyridine [2, 4] | Methyl-pyrimidine [2, 4] |
| 1H-Pyrrole [2, 4, 6, 7, 8] | 2-Amino-pyridine [2] | 2-Methyl-pyrimidine [7] |
| Imidazole [2] | 2-Cyano-pyridine [2] | 4-Methyl-pyrimidine [7] |
| Methyl-1H-pyrrole [2, 4, 8(2 isomers)] | 2-ethyl-pyridine [2, 4] | 2-Amino-pyrimidine [2] |
| Dimethyl-1H-pyrrole [2, 8(6 isomers)] | Dimethyl-pyridine [2] | Dimethylpyrimidine [2] |
| Ethyl-1H-pyrrole [2, 4] | Amino-methyl-pyridine [2] | Aminomethyl-pyrimidine [2] |
| Ethyl-methyl-1H-pyrrole [8(4 isomers)] | Trimethyl-pyridine [2] | Triazine [2] |
| Trimethyl-1H-pyrrole [2, 8(3 isomers)] | Dimethyl-amino-pyridine [2] | Indole [1] |
| Ethyl-dimethyl-1H-pyrrole [8(3 isomers)] | Pyridazine [2] | Benzimidazole [2] |
| Tetramethyl-1H-pyrrole [8] | Pyrazine [2, 4] | Methyl-benzimidazole [2] |
| Ethyl-trimethyl-1H-pyrrole [8(2 isomers)] | 2-Methyl-pyrazine [2, 4] | Indazole [2] |
| Diethyl-methyl-1H-pyrrole [8] | Dimethyl-pyrazine [2] | Ethyl-indole [2] |
| 3-Methyl-pyrrolidine [2] | 2,5-Dimethyl-3-ethyl-pyrazine [2] | Adenine [2] |
| **ISOMERS** | | |
| $C_4H_8$ [15 (3 isomers), 17] | $C_4H_7N$ [17] | $C_6H_{10}$ [13] |
| $C_4H_6$ [12, 15, 17] | $C_5H_{10}$ [13] | $C_7H_{12}$ [13] |
| $C_4H_5N$ [14, 15, 17] | $C_5H_8$ [13, 14, 17] | $C_8H_{10}$ [9] |
| **O-Compounds** | | |
| Carbon monoxide [5, 12, 17] | Acetic acid [1, 4] | Butanone [7] |
| Carbon dioxide [1, 4, 7, 11, 12, 13, 14, 15, 16, 17] | Acetamide [4, 7] | Fulminic acid [10] |
| Water [7, 12, 13, 14, 15, 16, 17] | Isocyanic acid [4, 11, 12] | Nitric oxide [11, 17] |
| Formamide [7, 10] | | |

*Table 2 List of compounds detected in previous thermal degradation and molecular composition studies of tholins. The superscripted numbers refer to the studies where each molecule has been detected, as listed in table 1, first column. When several isomers of a molecules have been detected, the number is specified in brackets.*

## II Experimental section

1. Samples synthesis with the PAMPRE laboratory experiment

PAMPRE (French acronym for Production d'Aérosols en Microgravité par Plasmas REactifs) is a laboratory Radio-Frequency cold plasma experiment developed to produce and analyze Titan's aerosol analogues. A detailed description of the PAMPRE set-up is found in (Szopa et al., 2006).
Three different tholin samples are produced at room temperature for this study. The syntheses are done under a 55 sccm continuous flow of a $N_2$:$CH_4$ gas mixture including 2%, 5% and 10% of methane. The reaction chamber is supplied by two high purity gas bottles, one of pure $N_2$ ($H_2O$ < 3 ppm, $O_2$ < 2 ppmv, $C_xH_y$ < 0.5 ppmv) and one containing a $N_2$–$CH_4$ mixture with 10% ± 0.2%v $CH_4$ ($H_2O$ < 3 ppmv, $O_2$ < 2 ppmv).
The three samples are named 2%-tholins, 5%-tholins and 10%-tholins respectively. The pressure in the reactor chamber is set to 0.86 mbar, corresponding to a quasi-static state of the gas. The injected RF power supplied by the generator is 30 W.
We use a $CH_4$ concentration range as wide as possible, since the methane ratio in Titan's atmosphere varies with altitude, but also with geological time scale (Tobie et al., 2009).
Tholins are trapped in a glass vessel, which is extracted at ambient air after opening the reactor. Samples are transferred from the glass vessel to tightly sealed vials, which are stored at room temperature (from one to a few dozen days) before analysis. A few hundred milligrams of 2%-tholins and 5%-tholins and a few dozen milligrams of 10%-tholins are produced for this study.

2. Pyrolysis – Gas Chromatography – Mass Spectrometry
   a) Apparatus

Pyr-GCMS experiments are carried out with a commercial gas chromatograph (GC) Trace GC Ultra and a commercial ion trap mass spectrometer (MS) ITQ 900, both from Thermo Scientific.
The GC is first equipped with a capillary MXT – Q PLOT column (30m-long, 0.25-mm inner diameter, 10 µm film thickness) from Restek. Its non-polar stationary phase of divinylbenzene is adapted for the analyses of molecules with one to three carbon atoms ($C_1$ to $C_3$) and for alkanes up to C12. Then the GC is equipped with a capillary Agilent DB-5MS column (30-m long, 0.25-mm inner diameter, 0.25 µm film thickness) with a non-polar stationary phase of (dimethyl-phenyl) siloxane polymer, particularly suited to polycyclic aromatic compounds and hydrocarbons analysis. These two different columns are chosen to allow the detection of compounds of a wide range of masses. The first one (Q-PLOT) targets the lightest organic and inorganic molecules, whereas the second one (DB-5MS) is rather dedicated to organic species of higher molecular weights.

A Pyrojector II pyrolysis oven (SGE Analytical Science, Trajan Scientific, Australia) is coupled to the GC Split-Splitless (SSL) injector. The pyrolyzer is chosen so that the oven can be heated to any temperature between room temperature and 900°C. The oven is flowed with helium at a constant rate with an adjustable pressure. With this set-up, a quasi-instantaneous pyrolysis is obtained when the sample is introduced into the oven. Pyrolysis products are transferred toward the head of the chromatographic column when an overpressure is maintained into the pyrolyzer compared to the chromatograph.

b) Analysis protocol

The sample is introduced into a 40-mm long, 0.5-mm outer diameter quartz liner. The liner is filled up with 1.0 ± 0.3 mg of tholins. Samples (2%, 5% and 10%-tholins) are pyrolyzed at each temperature between 200°C and 600°C with a 100°C step, in order to determine if the pyr-GCMS technique allows to discriminate between the 2%, 5% and 10%-tholins samples through their thermal decomposition products. A new un-pyrolyzed sample is used for each pyrolysis temperature. In order to ensure the cleanliness of the signal, a blank is recorded before each analysis using the same analytical conditions as for the samples.

The liner is introduced into the oven settled on a given pyrolysis temperature. The helium pressure is set to 23.8 psi into the pyrolyzer (5 psi overpressure with regard to the GC). The helium flow in the chromatographic column is 1.5 $ml.min^{-1}$ (129 kPa, 18.8 psi at 60°C). The split ratio is 1:10. The acquisition starts simultaneously with the injection.

The Q-PLOT column temperature program is: 60°C starting temperature, then a heating at a 10°$C.min^{-1}$ rate, and a final temperature of 190°C held for 17 min. The total GC run duration is 30 min. When the DB-5MS column is used, the analytical program is a 40°C isotherm for the 30 min of acquisition. Mass spectra are recorded in the 12-300 m/z range. The MS ion source temperature is 200°C. Identification of the pyrolysates is made by comparing their mass spectra with the NIST data base spectra and, for some of them, confirmed by injecting analytical standards.

Numerous chromatograms are run in order to determine the optimal analytical parameters, allowing the best chromatographic separation with a maximum number of detectable compounds. The systematic study of the effect of the pyrolysis temperature is carried out with the Q-PLOT column. Considering that the largest number of molecules analyzed with the Q-PLOT column is obtained at 600°C pyrolysis temperature, we focus the analysis done with the DB-5MS column on this specific temperature. The choice of the chromatographic column and the pyrolysis temperature is discussed in the next paragraphs.

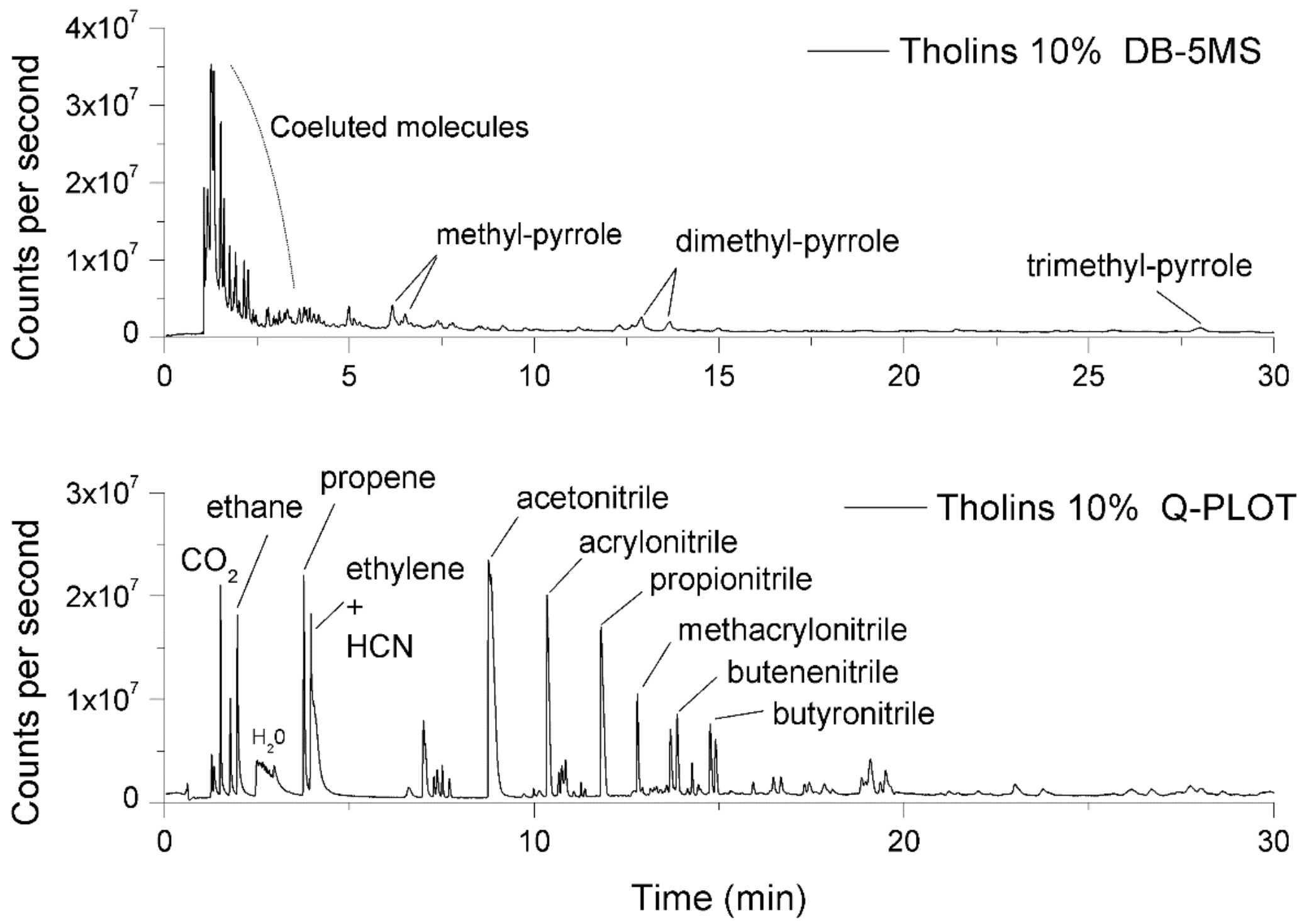

*Figure 1 Chromatograms obtained after a 600°C pyrolysis of 10%-tholins, with the DB-5MS column (top) and Q-PLOT column (bottom).*

## III Results and discussion

1. Species identified by pyr-GCMS

All the compounds identified after pyrolyses of 10%-tholins are listed in table 3. For the other samples, the compounds are reported in tables 5 and 6 as supplementary data. The pyrolysis temperature range where each molecule is detected is specified in the third sub-column. Molecules marked with an asterisk have been confirmed by injecting analytical standards.

No organic oxygen-bearing molecule was detected in a large amount in our analyses. Such molecules are expected to be present in small amount in Titan's aerosols. In spite of their low abundance, they are actually of strong interest in the frame of astrobiology. They are also known to be present in PAMPRE tholins as oxygen represents 2% of the atoms in the tholins (elemental analysis by (Fleury et al., 2014)). Oxygen-bearing molecules were also previously identified through GCMS techniques using an additional specific sample preparation described in Horst et al (2012) that involve liquid solvent extraction and derivatization. These oxygen bearing molecules are necessarily present in the chromatograms, but their detection can be difficult compared to the other molecules for different reasons: their signal can be under the detection limits allowed by this analysis; a few of them can be oxidized or decarboxylated (the abundant $CO_2$ detected being a marker of these processes)…

In order to analyze these specific minor oxygenated molecules in the context of Titan, the response of pyr-GCMS analysis towards these species can be enhanced by the addition of sample treatments (solvent extraction followed by chemical derivatization, as done in Hörst et al. 2012 for instance). Other analytical techniques, which are more sensitive to oxygen bearing species, can also be considered such as High Resolution Mass Spectrometry (Pernot et al. 2010).

Fig. 1 shows typical chromatograms obtained after a 600°C pyrolysis of 10%-tholins with the two columns. As expected, the lighter species are actually better separated with the Q-PLOT column. Nevertheless, nearly a third of the identified molecules are detected with the DB-5MS column only, and are highlighted in red in table 3. They are mainly N-heterocycles, including numerous pyrrole derivatives. The two complementary columns are thus essential to a comprehensive characterization of the pyrolysates.

**10% Q-PLOT et DB-5MS : 91 compounds**

| **ALKANES (4)** | | |
|---|---|---|
| **Empirical formula** | **Compound** | **T (°C)** |
| $C_2H_6$ | Ethane | 400-600 |
| $C_3H_8$ | Propane (co-elution ethylene) | 500-600 |
| $C_4H_{10}$ | Isobutane | 400-600 |
| $C_5H_{12}$ | 2-Methylbutane* | 500-600 |
| **ALKENES (6)** | | |
| **Empirical formula** | **Compound** | **T (°C)** |
| $C_3H_6$ | Propylene | 300-600 |
| $C_2H_4$ | Ethylene | 200-600 |
| $C_4H_8$ | 2-Methyl-propene | 300-600 |
| $C_4H_8$ | 2-Butene | 400-600 |
| $C_4H_9$ | 1-Butene | 400-600 |
| $C_5H_8$ | Cyclopentene* | 500-600 |
| **ALIPHATIC NITRILES (16)** | | |
| **Empirical formula** | **Compound** | **T (°C)** |
| HCN | Hydrogen cyanide | 200-600 |
| $C_2H_3N$ | Acetonitrile* | 200-600 |
| $C_3H_3N$ | Acrylonitrile | 200-600 |
| $C_3H_5N$ | Propionitrile | 200-600 |
| $C_4H_5N$ | Methacrylonitrile | 300-600 |
| $C_4H_7N$ | Isobutyronitrile* | 200-600 |
| $C_4H_5N$ | 2-Butenenitrile* (2 isomers) | 200-600 |
| $C_4H_5N$ | 3-Butenenitrile | 400-600 |
| $C_4H_7N$ | Butyronitrile* | 200-600 |
| $C_5H_7N$ | 2-Pentenenitrile | 300-600 |
| $C_5H_9N$ | Pentanenitrile* | 300-600 |
| $C_6H_{11}N$ | Hexanenitrile | 400-600 |
| $C_5H_5N$ | 2,4-Pentadienenitrile | 600 |
| $C_5H_7N$ | 3-Pentenenitrile* | 600 |
| $C_5H_7N$ | Methyl-butenenitrile | 600 |
| **ISOMERS (25)** | | |
| | **Compound** | **T (°C)** |
| | $C_4H_8$ | 500-600 |
| | $C_4H_6$ | 500-600 |
| | $C_5H_{10}$ (2 isomers) | 500-600 |
| | $C_5H_8$ (2 isomers) | 500-600 |

| **NITROGENOUS RINGS (24)** | | |
|---|---|---|
| **Empirical formula** | **Compound** | **T (°C)** |
| $C_4H_5N$ | Pyrrole | 300-600 |
| $C_4H_4N_2$ | Pyrimidine* | 400-600 |
| $C_5H_5N$ | Pyridine | 300-600 |
| $C_5H_6N_2$ | 2-Methylpyrimidine | 400-600 |
| $C_6H_7N$ | 2-Methylpyridine* (2-picoline) | 300-600 |
| $C_6H_7N$ | 3-Methylpyridine * (3-picoline) | 300-600 |
| $C_5H_6N_2$ | Amino-pyridine | 300-600 |
| $C_3H_5N_3$ | Methyl-1,2,4-triazole | 600 |
| $C_5H_7N$ | Methyl-pyrrole (**2** isomers) | 600 |
| $C_7H_9N$ | Dimethyl-pyridine (**2** isomers) | 600 |
| $C_6H_8N_2$ | Dimethyl-pyrimidine | 600 |
| $C_6H_9N$ | Dimethyl-pyrrole (**4** ismers) | 600 |
| $C_6H_9N$ | 3-Ethyl-pyrrole | 600 |
| $C_7H_{11}N$ | Ethyl-methyl-pyrrole (**4** isomers) | 600 |
| $C_7H_{10}N_2$ | Trimethyl-pyrazine | 600 |
| $C_7H_{11}N$ | Trimethyl-pyrrole | 600 |
| **CYCLES/AROMATICS (8)** | | |
| **Empirical formula** | **Compound** | **T (°C)** |
| $C_6H_6$ | Benzene | 400-600 |
| $C_8H_{10}$ | 1,2-Dimethyl-benzene* (o-xylene) | 400-600 |
| $C_8H_{10}$ | 1,4-Dimethyl-benzene* (p-xylène) | 400-600 |
| $C_7H_8$ | Toluene | 600 |
| $C_8H_{10}$ | Ethyl-benzene | 600 |
| $C_8H_{10}$ | Dimethyl-benzene (xylene) | 600 |
| $C_7H_5N$ | Benzonitrile | 600 |
| $C_9H_{12}$ | Ethyl-methyl-benzene | 600 |
| **To be confirmed (8)** | | |
| **Empirical formula** | **Compound** | **T (°C)** |
| $NH_3$ | Ammonia (co-elution $H_2O$) | 300-400 |
| $C_4H_6$ | Methylene-cyclopropane | 500-600 |
| $C_3H_6$ | Cyclopropane (co-elution acetonitrile) | 300-600 |
| $C_4H_9N$ | Amino-cyclobutane | 200-600 |
| $C_7H_{13}N$ | 2-Propyn-1-amine | 300-600 |

| | | | | | |
|---|---|---|---|---|---|
| | $C_6H_{10}$ (3 isomers) | 500-600 | $C_6H_9N$ | 3-Methyl-2-methylene-butanenitrile | 600 |
| | $C_6H_{12}$ (2 isomers) | 400-600 | $C_7H_9N$ | Benzylamine | 600 |
| | $C_5H_7N$ (7 **+1** isomers) | 300-600 | $C_9H_{12}$ | Trimethyl-benzene | 600 |
| | $C_5H_7N$ | 200-600 | | **Others** | |
| | $C_7H_{10}$ | 600 | **Empirical formula** | **Compound** | **T (°C)** |
| | $C_8H_{12}$ (3 isomers) | 600 | | Air | 200-600 |
| | $C_6H_7N$ | 600 | $N_2$ | Nitrogen | 400-600 |
| | | | $CO_2$ | Carbon dioxide | 200-600 |
| | | | $H_2O$ | Water | 200-600 |

*Table 3 Compounds identified after pyrolyses of 10%-tholins. Temperature range where the molecule has been detected are given in the third column. Molecules in red were detected with the DB-5MS column only. Molecules marked with an asterisk were confirmed by injecting analytical standards.*

Five main families of molecules are detected and identified in Table 3: alkanes, alkenes/cycles, aliphatic nitriles, aromatic and nitrogenous cycles. Alkanes, alkenes and nitriles have been generally detected in all the studies.

We notice the absence of alkynes compared to the work by (Coll et al., 1998) where dozen of alkynes molecules are detected. This difference is difficult to explain as tholins are produced in cold plasmas in both cases. A solid NMR study of (Derenne et al., 2012) is performed on tholins produced with the same experience as in our work. It shows no evidence of alkyne chemical functions in the bulk sample either, confirming the absence (or at least the low amount) of alkyne groups in our case.

Among the aromatic hydrocarbons, benzene and its derivatives are the most represented ones. But the pyr-GCMS studies also show that tholins' main structure is far from a Polycyclic Aromatic Hydrocarbon structure, in agreement with (Derenne et al., 2012).

Studies disagree regarding the number and the nature of N-heterocycles. For instance, (McGuigan et al., 2006) report no less than 23 pyrroles and alkyl-pyrroles, whereas other studies do not detected any of them. Several structures are only detected in (Khare et al., 1984): N-heterocycles with a bicyclic structure (indole, indazole), unstable cycles (aziridine, azetidine) and a few other molecules (adenine, imidazole, pyrrolidine). A possible pyrolysis temperature effect, which is further discussed, could be responsible for this presence of numerous N-heterocycles.

Linear amines are expected to be present in Titan aerosols (Raulin et al., 2012). However neither previous pyr-GCMS studies nor ours detect any of them in Titan's aerosol analogues. Actually, the GCMS technique has limitations for the detection of certain functional groups, including amines and other very polar molecules or molecules with a high molecular weight. Amines are known to adsorb on stainless steel present in the GCMS transfer lines. This is why chemical derivatization methods are under development in order to assist the pyr-GCMS detection (Buch et al., 2009; Freissinet et al., 2010; Geffroy-Rodier et al., 2009; Hörst et al., 2012). Other specific analyses involving different techniques (Infra-Red spectroscopy (Mahjoub et al., 2012), capillary electrophoresis (Cable et al., 2014), NMR (He et al., 2012), derivatization (Hörst et al., 2012)) are also carried out to detect amine molecules. These complementary techniques have actually confirmed the presence of primary and secondary amines in cold plasma tholins.

From these results, Pyr-GCMS is expected to enable the identification of numerous chemical signatures in Titan's aerosols. Two limitations are however highlighted that would require complementary analysis: the aromatic content could be overestimated when on the contrary amine molecules would possibly be underestimated.

2. Pyrolysis temperature

Tests have been done at 100°C and 150°C with the 5%-tholins. $CO_2$ and $H_2O$ are the major species detected with GCMS analysis in this temperature range, with no elution of organics. These molecules are not present in the reactive medium but come from the adsorption on the tholins' surface and are released for this range of pyrolysis temperatures. Previous studies of tholins' thermal degradation by thermogravimetric analysis (TGA) show indeed that between 80 and 200°C, the mass loss is mainly due to moisture vaporization ((He et al., 2014), (Nna-Mvondo et al., 2013)).

For the three tholins samples, thermal degradation starts above 150°C. At 150°C, acetonitrile becomes detectable. And above 200°C, light molecules – bearing up to four carbon atoms – are efficiently released.

The three samples follow the same evolution, with a higher total number of released species for the 10%-tholins sample. The evolution as a function of the pyrolysis temperature of the 10%-tholins pyrogram is given on Fig. 2. It illustrates the increase of the number and of the intensity of the chromatographic peaks when the pyrolysis temperature increases. This increase is observed up to 600°C where a maximum number of species are detected. The above-mentioned TGA studies highlight an important decomposition process in the 200-550°C temperature range. Above 600°C, carbonization becomes a major process, leading to the production of a more and more amorphous graphitic carbon nitride structure (Bonnet et al., 2015).

A 700°C pyrolysis test is done with a 10%-tholins sample: no additional identification of compounds is provided compared to the 600°C pyrolysis. These results differ from (Khare et al., 1984) who carried out sequential (a unique sample treated successively at different temperatures) and non-sequential pyrolyses up to 700°C. They reported a maximal pyrolysates richness at 300°C. The discrepancy could be explained by their use of a sequential process. Indeed we have observed that in case of sequential pyrolyses, the successive thermal treatments can lead to a faster degradation of the organic samples.

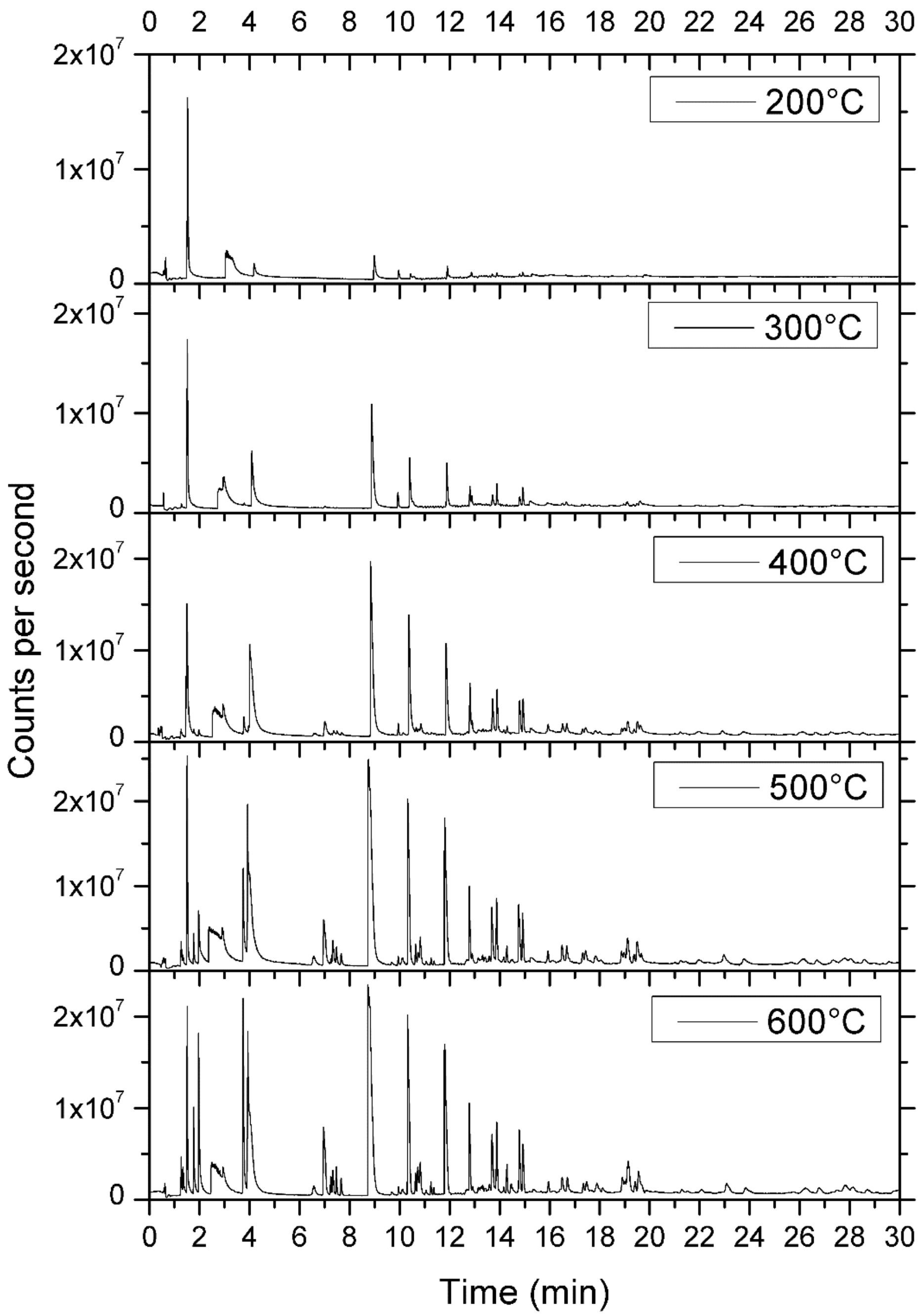

*Figure 2 Evolution of the 10%-tholins pyrogram with the pyrolysis temperature (Q-PLOT column).*

Figure 3 shows the evolution of the nature of the pyrolysates analyzed by both columns (Q-PLOT and DB-5MS) for the three tholins samples as a function of the pyrolysis temperature. Three major chemical families are taken into account: nitriles (linear and branched aliphatic nitriles and one aromatic nitrile, benzonitrile), hydrocarbons (linear and branched aliphatic hydrocarbons, cyclic and aromatic hydrocarbons together) and N-heterocycles.

The general trend is an increase of the number of the detected molecules when the temperature increases. The distribution among the three families is stable with the temperature for the 2%-tholins and 5%-tholins, but changes drastically for the 10%-tholins. This evolution is detailed below for each chemical family.

The number of hydrocarbons increases almost linearly with the temperature, but exceeds the number of nitriles only for the 10%-tholins in the 500-600 °C temperature range.

Except for acetonitrile, the release of nitriles starts at 200°C. Nitriles are by far the most numerous compounds detected at lower temperatures (200°C to 400°C) and are still the main released compounds at higher temperatures (400°C to 600°C) for the 2% and 5% tholins. The number of nitriles and N-heterocycles, detected when using the Q-PLOT column, is stable between 400°C and 500°C. Some heavier nitriles are detected at 600°C with the DB-5MS column.

Almost all the detected compounds are gaseous or liquid at room temperature. They result most likely from thermal degradation of heavier molecules in the solid grains. This applies for both nitriles (liquid at room temperature), and small hydrocarbons such as ethane or propane, which are observed only at 400°C and above.

Compounds detected at lower temperatures (200°C, 300°C) are the less thermally evolved molecules released from the samples and then certainly the most representative of the original material. They are mainly nitriles and ethylene, along with small-unsaturated hydrocarbons (up to 4 carbon atoms) for tholins synthetized from a 10%-$CH_4$ mixture.

Some of the compounds that have been identified in the tholins pyrolysates are also detected by GC-MS in the gas phase during the tholins synthesis (Gautier et al., 2011), especially (i) hydrocarbons: ethane, ethylene, propane, propylene, isobutene and (ii) nitriles: HCN, acetonitrile, propionitrile, acrylonitrile, butyronitrile, isobutyronitrile, pentanenitrile. Only three aromatics are detected both in tholins and in the gas phase: pyrrole, benzene and triazine. All these species are moreover potential precursors of numerous other detected molecules found as their alkyl derivatives.

N-heterocycles are mainly detected at higher temperatures in all the samples. The numerous N-heterocycles and aromatic hydrocarbons detected between 400°C and 600°C are mostly formed through cyclization and aromatization reactions at high temperatures (Bergman, 1973; John and Tour, 1994; Lockhart et al., 1981), including cyclization of nitrile groups (Devasia et al., 2003). A large proportion of N-heterocycles are pyrrole and its methyl- and ethyl- derivatives. This result is consistent with the study of (McGuigan et al., 2006), who find that alkyl-pyrroles are among the most dominant peaks. As these species are here more likely detected at high temperatures, they may mainly be due to the pyrolysis. The contribution of pyrrolic cycles in the tholins structure have therefore to be carefully considered in the study of (McGuigan et al., 2006). Nevertheless pyrrole and triazine have previously been identified in the gas phase and in tholins after derivatization (Hörst et al., 2012). A few N-heterocycles are therefore also present as molecular components in the samples and can contribute as a moderate direct release source during the pyrolysis process.

This study confirms that the pyrolysis temperature of 600°C used by ACP is actually an optimum for the detection of products from Titan's aerosols, but that secondary productions of aromatics and N-aromatics are also expected at this temperature. If detected, aromatic molecules can therefore not be firmly considered as representative structures in Titan's aerosols by this technique.

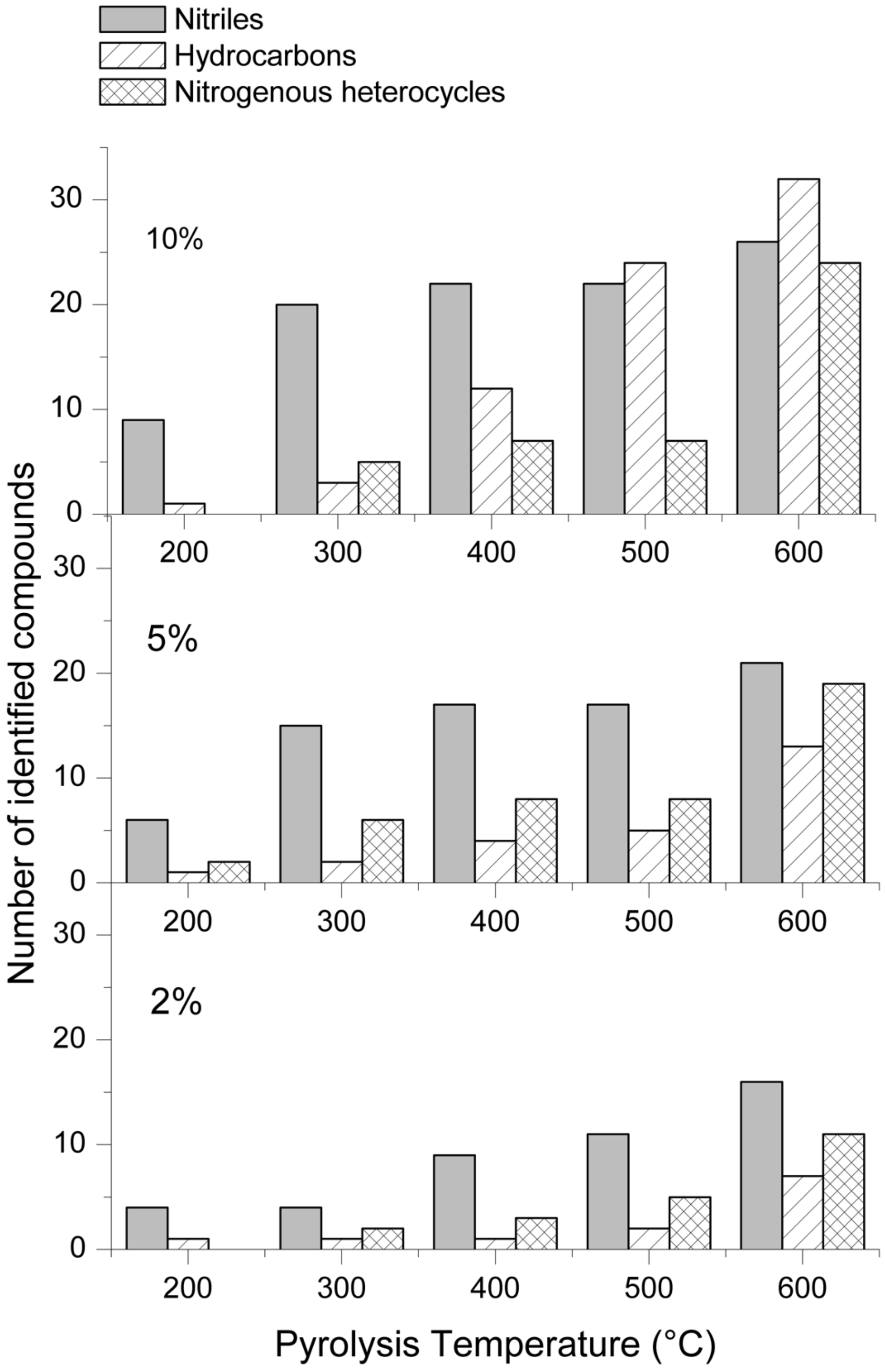

*Figure 3 Evolution of the number of types of molecules grouped by chemical families: nitriles (aliphatic and aromatic together), hydrocarbon (aliphatic, aromatic and cyclic together) and N-heterocycles as a function of the pyrolysis temperature for the three tholins samples: 10% (top), 5% (middle), 2% (bottom), (detected with Q-PLOT and DB-5MS columns).*

3. Influence of $CH_4$ concentration in the initial gas mixture on the tholins molecular structuration

Fig. 4 shows the chromatograms obtained after pyrolysis of the three samples at 400°C with the Q-Plot column. All the three chromatograms have the same major pattern between 8 min and 15 min. This pattern contains the most intense peaks and is resulting from mainly nitriles (see Fig. 1). Another major peak is ethylene, eluted at about 4 min.
The main difference between the three samples is the number of molecules generated by the pyrolysis and detected by both columns. 40, 62 and 91 different compounds are identified in the pyrograms of the 2%, 5% and 10%-tholins respectively. The number of pyrolysates increases when the $CH_4$ ratio increases for whatever chemical family (Fig. 5).

Table 4 provides the number of compounds of each family found in the three samples at 600°C. The 2% and 5%-tholins are rather similar, with however a nitriles/hydrocarbons ratio slightly greater for the 2%-tholins. The 10%-tholins sample differs with a hydrocarbon percentage nearly twice as large as in the two other samples. This higher number of hydrocarbons is at the expense of nitriles. This is consistent with a higher content of hydrocarbon chains in the molecular structure of tholins. Other studies confirm this structure. An increase of the C/N ratio according to the initial methane concentration is consistently reported in (Sciamma-O'Brien et al., 2010). A multiplication of the methylene ($-CH_2-$) pattern in the 10%-tholins is found in (Pernot et al., 2010). Infrared absorption analyses show an increase of aliphatic C-H groups when the methane ratio increases (Gautier et al., 2012). The important aliphaticity of the 10%-tholins may make the material more thermolabile. This fragility would explain the increase in the number of pyrolysates and the higher signal obtained for each compound released by this sample in comparison to the other tholins samples.

| **Sample** | **Nitriles** | | **Hydrocarbons** | | **Nitrogenous heterocycles** | | **Total number** |
|---|---|---|---|---|---|---|---|
| | Number | Relative | Number | Relative | Number | Relative | |
| **2%** | 13 | 43 % | 6 | 20 % | 11 | 37 % | 30 |
| **5%** | 16 | 35 % | 10 | 22 % | 19 | 42 % | 45 |
| **10%** | 17 | 26 % | 24 | 37 % | 24 | 37 % | 65 |

*Table 4 Absolute number and percentage of compounds for the three major chemical families regarding to the total number of compounds in the three families (last column) at 600°C.*

All the compounds identified among the 2% and 5%-tholins pyrolysates are also detected in the 10%-tholins pyrolysates, except two of them. The 1,3,5-triazine ($C_3H_3N_3$) is only detected in the 2%-tholins after a 300 to 600°C pyrolysis. And the amino-pyrazine ($C_4H_5N_3$) is detected in 2% and 5%-tholins pyrolysed between 400°C and 600°C. Only one compound bearing three nitrogen atoms is released by the 10% (and 5%) tholins, at 600°C: methyl-1,2,4-triazole. The formation of poly-nitrogen molecules seems therefore facilitated by low initial $CH_4$ concentrations.

Besides these results, a temperature program was also applied to the DB-5MS column (ramp up to 190°C) in order to allow the elution of potentially heavy compounds. The 10%-tholins mass spectra show a pattern consistent with oligomers (Fig. 6). This pattern is less distinct in the 5%-tholins and nearly absent in the 2%-tholins. We suspect that the quantity of alkyl chains increases with the methane ratio, and consequently, the probability of copolymerization. With this program, we also detect additional alkyl-pyrroles and three triazine derivatives: 2,4,6-triamino-1,3,5-triazine (melamine), 2,4-diamino-6-methyl-1,3,5-triazine and 2,4-diamino-1,3,5-triazine. The first two molecules are also detected by multidimensional NMR ((He and Smith, 2014), (Derenne et al., 2012)). Moreover, a UV-Raman study by (Quirico et al., 2008) on tholins produced with the PAMPRE experiment suggests that triazine and its derivatives are major compounds in the tholins molecular structure.

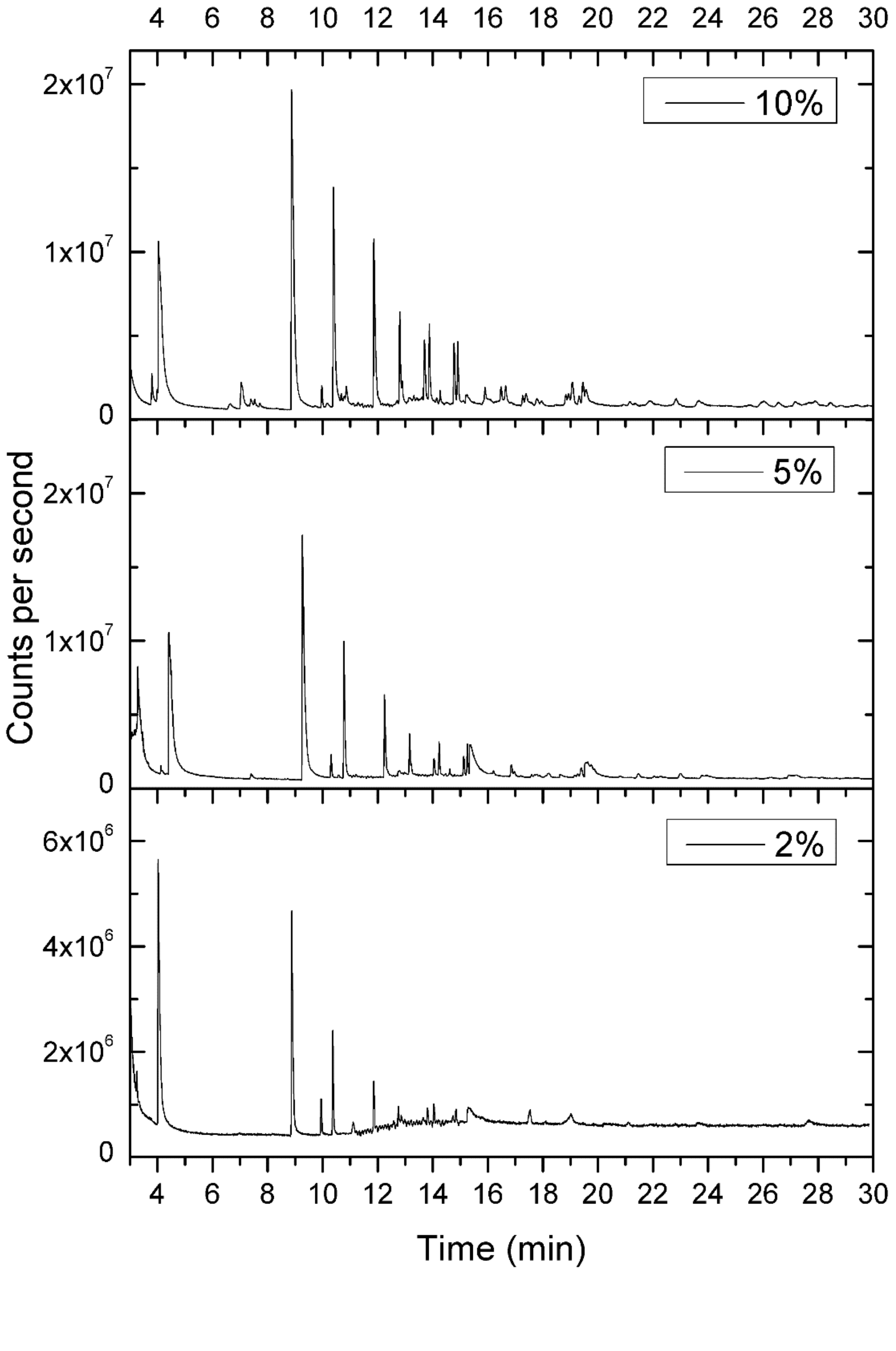

*Figure 4 Chromatograms obtained after a 400°C pyrolysis of 2%-tholins (bottom), 5%-tholins (middle) and 10%-tholins (top) with the Q-PLOT column.*

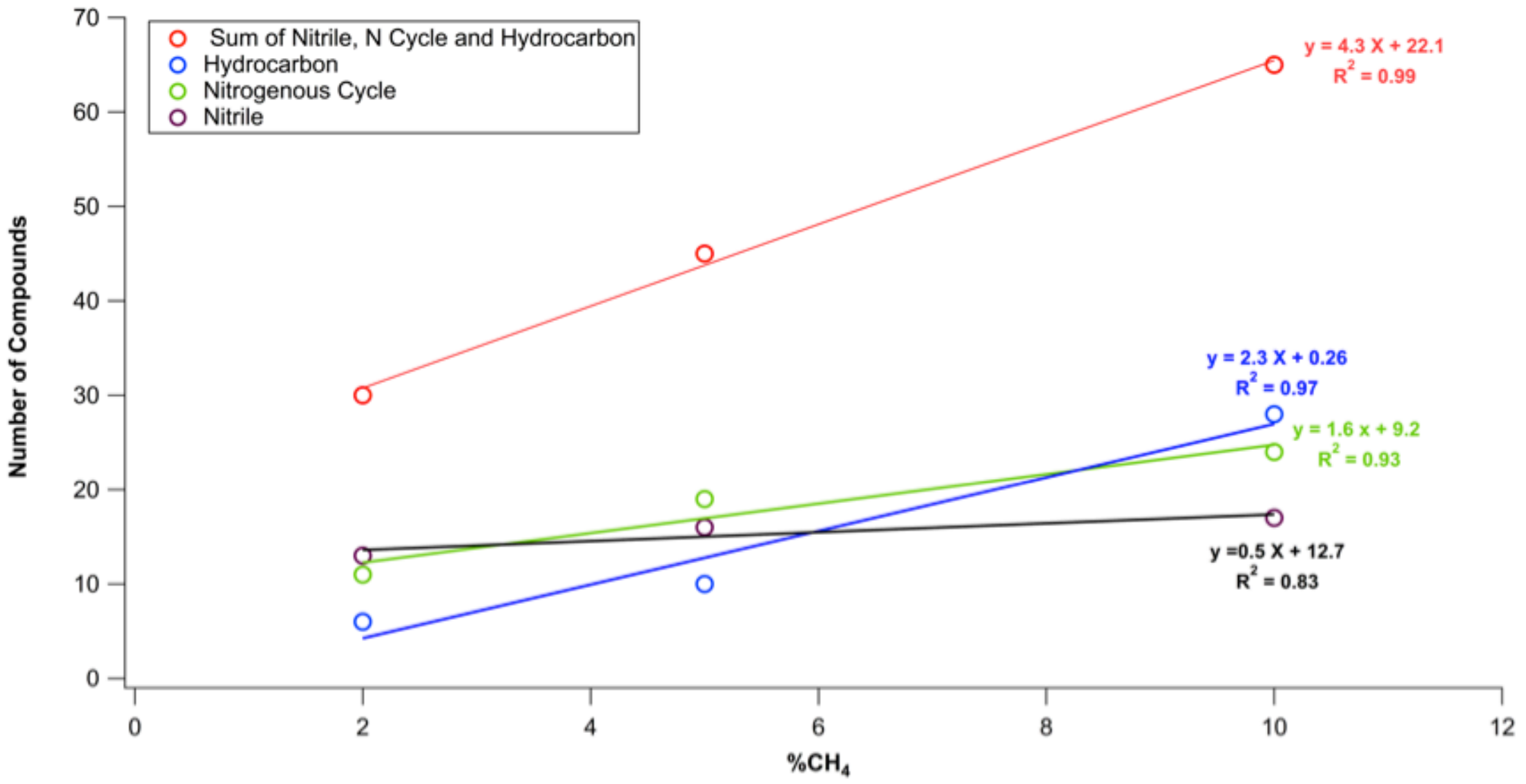

*Figure 5 Evolution of the number of compounds released in each chemical family versus %$CH_4$*

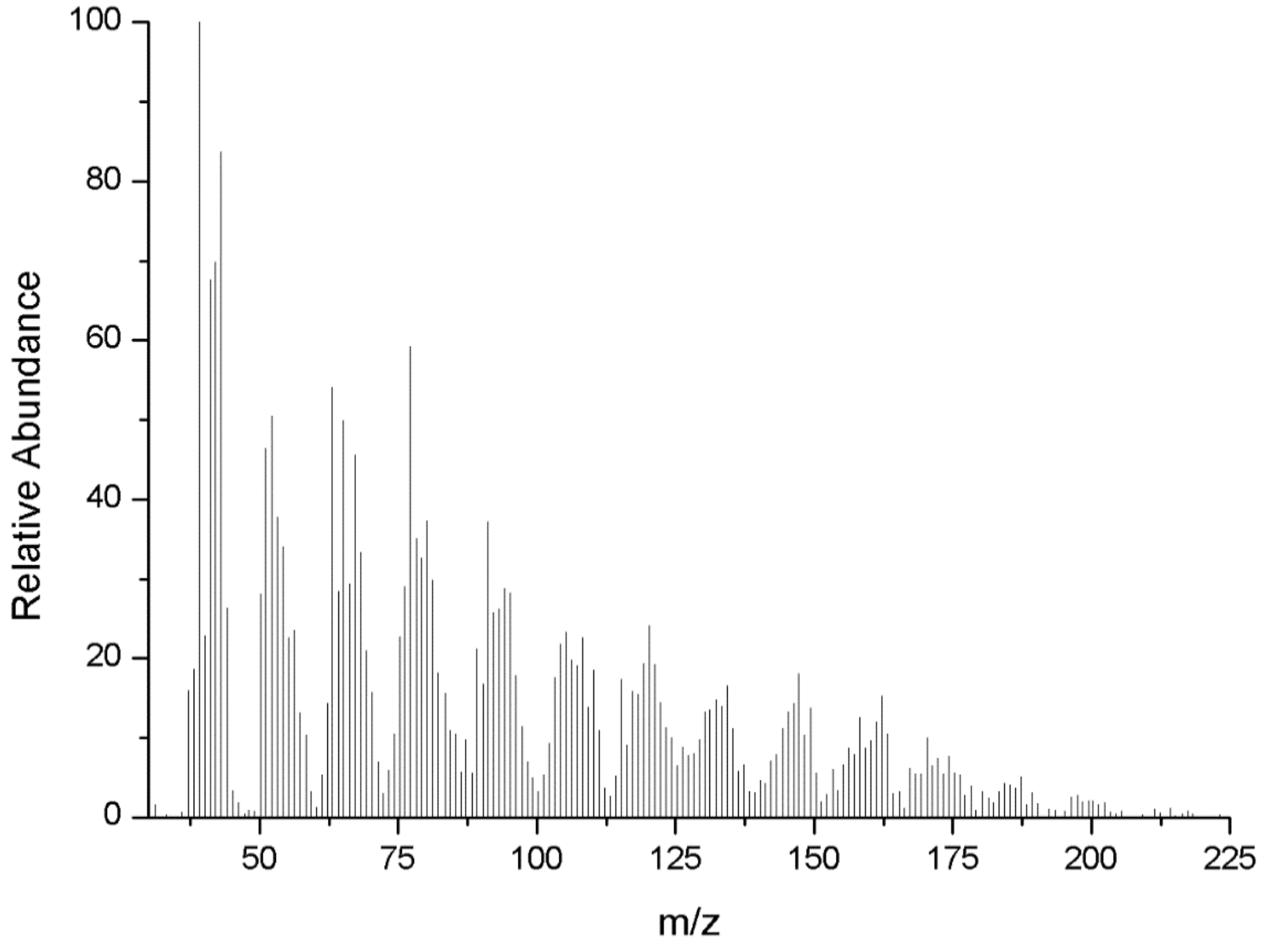

*Figure 6: Mass-pattern obtained after a 600°C pyrolysis of the 10%-tholins with the DB-5MS column.*

The evolution of tholins composition with methane percentage in the initial gas mixture is also consistent with previous analyses performed with Orbitrap High Resolution Mass Spectrometry on tholins produced with PAMPRE (Gautier et al., 2014). They similarly detect a large amount of nitrogen bearing molecules and show that the number of molecules with high C/N ratios depends on the amount of methane. The hetero-polymer pattern found in tholins by Orbitrap analysis is also much more pronounced in the 10%-tholins, just as with our pyr-GC-MS analysis (Fig. 6).

Pyr-GCMS signatures are different from a sample to another according to their chemical composition. Pyr-GCMS technique provides a chemical fingerprint of the analyzed organic material. The aerosols undergo physical and chemical evolution during their atmospheric descent in Titan's atmosphere. Our study shows that the Pyr-GCMS technique is sensitive to chemical evolution and would be a valuable diagnosis to probe such an evolution in Titan's atmosphere.

## 4. Semi-quantitative study

We perform a semi-quantitative study of nitriles, the main chemical families represented in the chromatograms. The chromatograms are reproducible – especially for the 10%-tholins – and the analyses are carried out within a short duration (2 days), so that there is no loss of sensitivity with the instruments. We can therefore compare the quantities of the different nitriles (in arbitrary units). Seven nitriles containing 2 to 4 carbon atoms with well-resolved chromatographic peaks are chosen: acetonitrile ($C_2H_3N$), acrylonitrile ($C_3H_3N$), propionitrile ($C_3H_5N$), methacrylonitrile ($C_4H_5N$), 2-and-3-butenentrile ($C_4H_5N$) and butyronitrile ($C_4H_7N$). Fig. 7 shows the evolution of each nitrile's peak area as a function of the pyrolysis temperature. The main uncertainty on the intensities comes from the sampling in the pyrolysis liners. The introduced mass varies by about 20% from sample to sample. The chromatographic peak intensities are directly affected by this variation of the sample mass, as confirmed when reproducing the experiment.

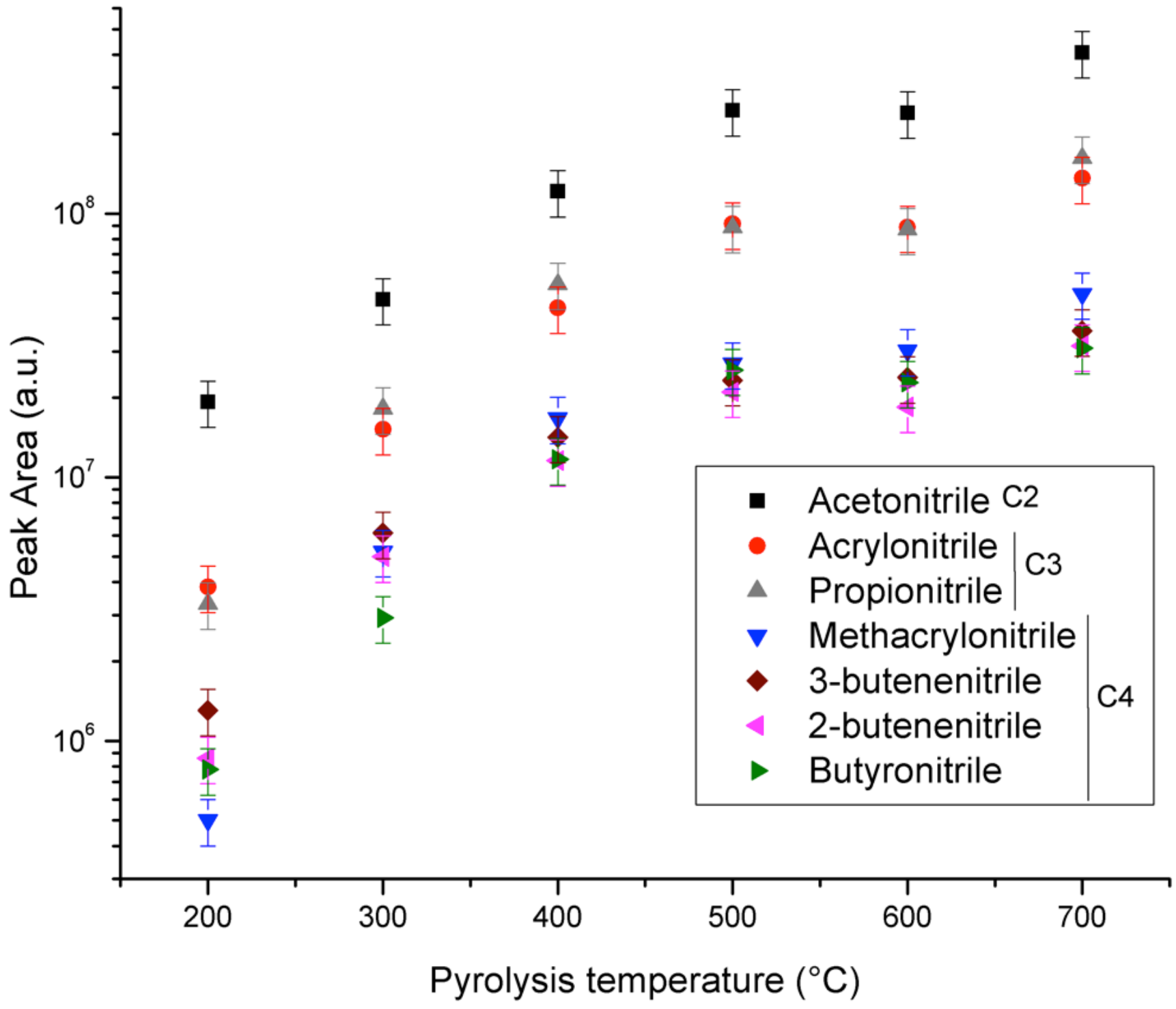

*Figure 7 Evolution of the peaks area with pyrolysis temperature for major nitriles identified in 10%-tholins pyrolysates.*

The overall trend is an increase of the quantity of detected molecules with the temperature. This is probably because the tholins core is more deeply degraded with a higher pyrolysis temperature. This is consistent with the TGA study performed by (He et al., 2014). Between 200°C and 700°C, TGA shows a continuous and nearly linear mass loss. Two main phases are observed. The release of nitriles increases up to 500°C, then it is stable, before increasing again beyond 650°C. These two steps correspond to the degradation processes described in He et al., 2014, with a main thermal degradation phase followed by a graphitization process.
All the selected nitriles show a behavior similar to propionitrile between 300°C and 700°C. The ratios are quite constant regardless of the temperature, showing that the material is homogenous in terms of nitriles distribution.

Among the chromatographic peaks areas, three groups can be distinguished: (i) the C2-nitrile, acetonitrile, is largely dominant; (ii) the C3-nitriles, acrylonitrile and propionitrile, are in intermediate quantities; (iii) and all the C4-nitriles are in lower proportions. As the ionization cross-section increases with the number of atoms in the molecules, this distribution is even emphasized when it is converted into quantity of released pyrolysates. This highlights a relation between the mass and the quantity of each nitrile extracted from tholins: the quantity increases with the length of the carbon chain. This behavior is, at least partially, explained by the thermal fragmentation of the longest chains, increasing the quantity of small nitriles. The important quantity of small nitriles observed in our tholins may also be explained by a greater quantity of precursors available in the gas phase. Indeed a relative quantification of nitriles as a function of their carbon numbers has been performed in the PAMPRE gas phase (Gautier et al., 2011). It shows an exponential decrease of the nitriles concentration when the number of carbon atoms increases.

The detection of HCN by the ACP instrument is in agreement with a significant contribution of nitriles in Titan's aerosols. The pyr-GCMS analysis performed here on Titan's aerosols analogues shows that in a future mission to Titan, the implementation of Pyr-GCMS would provide quantitative data on the nitrile content in Titan's aerosols.

**IV Conclusion**

The Pyr-GCMS technique was already successfully implemented for chemical analysis in several space missions: Viking 1 and 2 (Biemann et al., 1977), MSL on Mars (Mahaffy et al., 2012), Rosetta on comets (Rosenbauer et al., 2012). A first attempt was made to also analyze *in situ* the organic aerosols in Titan's atmosphere with the Aerosol Collector and Pyrolyzer (ACP) instrument onboard the Cassini-Huygens mission (Israel et al. 2005). Some technical issues during the descent in Titan's atmosphere prevented any extensive analysis, but the identification of the main pyrolysis fingerprint of the aerosols was made possible. It revealed a refractory nucleus releasing hydrogen cyanide, ammonia and C2 hydrocarbons during pyrolysis. As the aerosol chemical composition remains largely unknown, their *in situ* analysis remains an important issue for a future mission to Titan. In this frame, the pyr-GCMS technique could be used again as it is robust and already space qualified, but it is still to be improved and optimized in the case of Titan.

This is the aim of the present work. We proposd a comprehensive Pyr-GCMS study of interest for Titan using laboratory analogues of Titan's aerosols. The analogues are chosen as the most representative of the ACP signature (Coll et al., 2013). We explore two main parameters for the analysis: the pyrolysis temperature and the chemical composition of the analogues, controlled by the initial methane content during their synthesis. Three samples are synthetized with different methane ratios – namely 2%, 5% and 10% of $CH_4$ in $N_2$ – in order to compare their molecular structures. Moreover, we aim at detecting exhaustively the molecules released during the pyrolysis process by using two complementary chromatographic columns. It is also a first attempt to quantify the pyr-GC-MS results after an optimization of the analytical conditions.

Firstly, considering the effect of the pyrolysis temperature, the highest signal and the largest number of identified molecules are obtained after a 600°C pyrolysis with the Q-Plot column. The semi-quantitative study performed on nitriles shows that the quantity of released molecules (from the chromatographic peaks area) increases linearly with the temperature up to 500°C. Moreover, the ratios between nitrile quantities are constant at every pyrolysis temperatures, confirming a homogeneous distribution of nitriles in the material. However, the material becomes graphitized above 500°C. Additional cracking molecules and aromatization of the released species become more and more problematic for the interpretation of the results as the temperature increased. Our study points out the 500°C pyrolysis temperature as the best compromise for the detection of the maximum of species (number and quantities) with a minimum of pyrolysis artefacts.

Secondly, we highlight some variability both in the nature and in the quantity of the molecules released from the different chemical samples. The formation of poly-nitrogen compounds is facilitated by low $CH_4$ mixing ratios. The samples mainly differ by the number of released compounds, which increases according to the increase of the initial $CH_4$ concentration. This could be explained by the higher thermolability of the sample with increasing aliphaticity. The 10%-tholins sample is actually found to contain larger amounts of hydrocarbon chains.

Pyrolyses of the three samples lead to the same major pattern formed of nitriles and ethylene. Saturated small hydrocarbons (ethane and propane) massively appear at temperatures above 600°C suggesting that they correspond to cracking molecules. The nitrile fingerprint of our samples, in agreement with the HCN detection done with ACP in Titan's aerosols, suggests that if the ACP instrument had worked perfectly, it would have probably identified a similar pattern in the aerosols of Titan. The detection of $NH_3$ by ACP can however not be reproduced in our analysis due to amine losses in the MS transfer lines. The design of the ACP-GCMS coupling part was actually optimized to prevent this effect to be significant (Israel et al., 2002).

Finally, the use of two complementary columns enables to detect a hundred of molecules in the pyrolysates, including aliphatic and aromatic hydrocarbons, nitriles and nitrogenous heterocycles. The Q-Plot column is sufficient to highlight the major molecules, up to 7 heavy atoms (carbon and nitrogen). The DB-5ms column allows detecting larger molecules, up to 9 heavy atoms (numerous N-heterocycles), but also a polymeric mass-pattern showing the interest of pyrolysis to extract large structures representative of the material. Unfortunately, the analysis of this polymeric mass-pattern has been made difficult due to overlapping signatures from the co-elution of numerous fragments. Nevertheless, we are able to identify three triazine derivatives (melamine, 2,4-diamino-6-methyl-1,3,5-triazine and 2,4-diamino-1,3,5-triazine), which are known to be of major importance in the tholins bulk composition (Quirico et al., 2008). These two columns are therefore essential for a comprehensive characterization of the solid samples.

Moreover, numerous species detected here in PAMPRE tholins are qualitatively and quantitatively consistent with compounds identified previously in the PAMPRE gas phase and with NMR studies of the bulk tholins. This shows that pyr-GC-MS is a suitable analysis tool for the solid phase, since it does not lead to a complete alteration of the tholin's chemical composition. As previously suggested by (Gautier et al., 2011), this also implies that nitriles found in abundance in the gas phase are precursors for tholins, rather than being left out because they are not used during the tholins synthesis.

From these results, several conclusions can be drawn on the potential of Pyr-GCMS for a future mission to Titan. A pyrolysis temperature of 600°C used by ACP is an optimum for the detection of products composing Titan's aerosols, but aromatics and N-aromatics are expected to be overestimated at this temperature. As Pyr-GCMS is sensitive to chemical evolution, it would be a valuable diagnosis to probe a chemical evolution of Titan's aerosols in the atmosphere. And Pyr-GCMS would also provide a tool for quantitative analysis of the nitriles content in Titan's aerosols.

Considering the efficiency of the Pyr-GCMS technique in the context of space exploration, there is no doubt that it will be implemented in future space probes, as already planned for the MOMA-GCMS experiment on the ExoMars-2018 Martian rover (Goetz et al., 2011). As this technique is simple and robust it could also be proposed for a future mission to Titan to analyze the solid materials present in the atmosphere and on the surface. In the present work, we optimized the Pyr-GCMS technique focusing on a future mission to Titan, enabling to highlight the main nitrile pattern, extrapolating the promising results by the ACP instrument on Huygens. These results will also be of significant help for interpreting future Pyr-GCMS data of solid organic materials, not just in the case of Titan, but throughout the entire Solar System, from cometary materials, to planetary surfaces, through interplanetary dust particles and asteroids.

**Acknowledgements**

We gratefully acknowledge the Programme National de Planétologie (PNP, INSU/CNES, France) for support of this research. NC acknowledges the European Research Council for their financial support (ERC Starting Grant PRIMCHEM, grant agreement n°636829). We thank David Dubois for his help in the English improvement of the article.

## Supplementary data

**2% Q-Plot et DB-5ms : 40 compounds**

| **ALKANES (1)** | | |
|---|---|---|
| **Empirical formula** | **Compound** | **T (°C)** |
| $C_2H_6$ | ethane | 600 |

| **ALKENES (4)** | | |
|---|---|---|
| **Empirical formula** | **Compound** | **T (°C)** |
| $C_3H_6$ | propylene | 500-600 |
| $C_2H_4$ | ethylene | 200-600 |
| $C_4H_8$ | 2-methyl-propene | 600 |
| $C_5H_8$ | cyclopentene* | 600 |

| **ALIPHATIC NITRILES (13)** | | |
|---|---|---|
| **Empirical formula** | **Compound** | **T (°C)** |
| HCN | hydrogen cyanide (coelution ethylene) | 200-600 |
| $C_2H_3N$ | acetonitrile* | 200-600 |
| $C_3H_3N$ | acrylonitrile | 200-600 |
| $C_3H_5N$ | propionitrile | 200-600 |
| $C_4H_5N$ | methacrylonitrile | 400-600 |
| $C_4H_7N$ | isobutyronitrile* | 500-600 |
| $C_4H_5N$ | 2-butenenitrile* (2 isomers) | 400-600 |
| $C_4H_5N$ | butenenitrile | 600 |
| $C_4H_7N$ | butyronitrile* | 400-600 |
| $C_5H_7N$ | 2-methylene-butyronitrile | 600 |
| $C_5H_5N$ | 2,4-pentadienenitrile | 600 |
| $C_5H_7N$ | methyl-butenenitrile | 600 |

| **ISOMERS (4)** | | |
|---|---|---|
| | **Empirical formula** | **T (°C)** |
| | $C_4H_8$ | 600 |
| | $C_5H_7N$ | 600 |
| | $C_5H_7N$ | 500-600 |
| | $C_5H_7N$ | 400-600 |

| **NITROGENOUS RINGS (11)** | | |
|---|---|---|
| **Empirical formula** | **Compound** | **T (°C)** |
| $C_3H_3N_3$ | 1,3,5-triazine* | 300-600 |
| $C_4H_5N$ | pyrrole | 500-600 |
| $C_4H_4N_2$ | pyrimidine* | 300-600 |
| $C_4H_5N_3$ | aminopyrazine | 400-600 |
| $C_5H_5N$ | pyridine | 500-600 |
| $C_5H_6N_2$ | 2-methylpyrimidine | 600 |
| $C_6H_7N$ | 2-methylpyridine* | 600 |
| $C_6H_7N$ | 3-methylpyridine* | 600 |
| $C_5H_6N_2$ | amino-pyridine | 600 |
| $C_5H_7N$ | methyl-pyrrole (2 isomers) | 600 |

| **AROMATICS (1)** | | |
|---|---|---|
| **Empirical formula** | **Compound** | **T (°C)** |
| $C_6H_6$ | benzene | 600 |

| **To be confirmed (6)** | | |
|---|---|---|
| **Empirical formula** | **Compound** | **T (°C)** |
| $C_2H_2$ | acetylene | 600 |
| $NH_3$ | ammonia | 300-600 |
| $C_3H_6$ | cyclopropane (coelution acetonitrile) | 500-600 |
| $C_4H_9N$ | aminocyclobutane | 200-600 |
| $C_7H_{13}N$ | 2-propyn-1-amine | 600 |
| $C_7H_8$ | toluene | 500-600 |

| **Others** | | |
|---|---|---|
| **Empirical formula** | **Compound** | **T (°C)** |
| | air | 200-600 |
| $N_2$ | nitrogen | 200-600 |
| $CO_2$ | carbon dioxide | 200-600 |
| $H_2O$ | water | 300-600 |

*Table 5 Compounds identified after pyrolyses of 2%-tholins. Temperature range where the molecule has been detected are noted in the third column.*

**5% Q-Plot et DB-5ms : 62 compounds**

| ALKANES (3) | | |
|---|---|---|
| **Empirical formula** | **Compound** | **T (°C)** |
| $C_2H_6$ | ethane | 400-600 |
| $C_3H_8$ | propane | 600 |
| $C_4H_{10}$ | isobutane | 600 |

| ALKENES (4) | | |
|---|---|---|
| **Empirical formula** | **Compound** | **T (°C)** |
| $C_3H_6$ | propylene | 400-600 |
| $C_2H_4$ | ethylene | 200-600 |
| $C_4H_8$ | 2-methyl-propene | 300-600 |
| $C_5H_8$ | cyclopentene* | 600 |

| ALIPHATIC NITRILES (15) | | |
|---|---|---|
| **Empirical formula** | **Compound** | **T (°C)** |
| HCN | hydrogen cyanide (coelution ethylene) | 200-600 |
| $C_2H_3N$ | acetonitrile * | 200-600 |
| $C_3H_3N$ | acrylonitrile | 200-600 |
| $C_3H_5N$ | propionitrile | 200-600 |
| $C_4H_5N$ | methacrylonitrile | 300-600 |
| $C_4H_7N$ | isobutyronitrile* | 300-600 |
| $C_4H_5N$ | 2-butenenitrile* (2 isomers) | 200-600 |
| $C_4H_5N$ | 3-butenenitrile | 300-600 |
| $C_4H_7N$ | butyronitrile* | 300-600 |
| $C_5H_9N$ | pentanenitrile* | 400-600 |
| $C_5H_5N$ | 2,4-pentadienenitrile | 600 |
| $C_5H_7N$ | methyl-butenenitrile | 600 |
| $C_5H_7N$ | 3-pentenenitrile | 600 |

| ISOMERS (10) | | |
|---|---|---|
| | **Formule brute** | **T (°C)** |
| | $C_4H_8$ | 600 |
| | $C_4H_6$ | 600 |
| | $C_5H_7N$ (5 isomers) | 300-600 |
| | $C_5H_7N$ | 400 |
| | $C_7H_{10}$ (2 isomers) | 600 |

| Others | | |
|---|---|---|
| **Empirical formula** | **Compound** | **T (°C)** |
| | air | 200-600 |
| $N_2$ | nitrogen | 200-600 |
| $CO_2$ | carbon dioxide | 200-600 |
| $H_2O$ | water | 200-600 |

| NITROGENOUS RINGS (19) | | |
|---|---|---|
| **Empirical formula** | **Compound** | **T (°C)** |
| $C_4H_5N$ | pyrrole | 300-600 |
| $C_5H_7N$ | methyl-pyrrole | 400-600 |
| $C_5H_7N$ | methyl-pyrrole (2 isomers) | 600 |
| $C_4H_4N_2$ | pyrimidine* | 200-600 |
| $C_4H_5N_3$ | aminopyrazine | 300-400 |
| $C_5H_5N$ | pyridine | 200-600 |
| $C_6H_7N$ | 2-methylpyridine* | 300-600 |
| $C_6H_7N$ | 3-methylpyridine* | 300-600 |
| | | |
| $C_5H_6N_2$ | amino-pyridine | 400 |
| $C_3H_5N_3$ | methyl-triazole | 600 |
| $C_5H_7N$ | methyl-pyrrole (**2** isomers) | 600 |
| $C_6H_8N_2$ | dimethyl-pyrimidine | 600 |
| $C_6H_9N$ | dimethyl-pyrrole (**4** isomers) | 600 |
| $C_7H_{11}N$ | trimethyl-pyrrole | 600 |

| AROMATICS (3) | | |
|---|---|---|
| **Empirical formula** | **Compound** | **T (°C)** |
| $C_6H_6$ | benzene | 500-600 |
| $C_8H_{10}$ | 1,4-dimethyl-benzene* | 600 |
| $C_7H_5N$ | benzonitrile | 600 |

| To be confirmed (8) | | |
|---|---|---|
| **Empirical formula** | **Compound** | **T (°C)** |
| $C_2H_2$ | acetylene | 600 |
| $NH_3$ | ammonia (coelution $H_2O$) | 300-600 |
| $C_3H_6$ | cyclopropane (coelution acetonitrile) | 500-600 |
| $C_4H_8$ | butene | 600 |
| $C_4H_9N$ | aminocyclobutane | 200-600 |
| $C_7H_{13}N$ | 2-propyn-1-amine | 600 |
| $C_7H_8$ | toluene | 500-600 |
| $C_5H_{13}N$ | 2-pentanamine | 400-600 |

*Table 6 Compounds identified after pyrolyses of 5%-tholins. Temperature range where the molecule has been detected are noted in the third column.*